\documentclass[]{aastex631}

\usepackage{float}

\newcommand{\kR}{\mbox{$\kappa_{\rm R}$}}

\shorttitle{Opacities with Solid Grains}
\shortauthors{Paola Marigo et al.}
\graphicspath{{./}{figures/}}
\usepackage{multirow}
\usepackage{booktabs}
\usepackage{threeparttable}
\usepackage{courier}

\begin{document}

\title{\AE SOPUS 2.0: Low-Temperature Opacities with Solid Grains}

\correspondingauthor{Paola Marigo}
\email{paola.marigo@unipd.it}

\author[0000-0002-9137-0773]{Paola Marigo}
\affiliation{Department of Physics and Astronomy G. Galilei,
University of Padova, Vicolo dell'Osservatorio 3, I-35122, Padova, Italy}
\author[0000-0002-8900-3667]{Peter Woitke}
\affiliation{Space Research Institute, Austrian Academy of Sciences, Schmiedlstr. 6, 8042, Graz, Austria}
\author[0000-0001-5736-628X]{Emanuele Tognelli}
\affiliation{CEICO, Institute of Physics of the Czech Academy of Sciences, Na Slovance 2, 182 21 Praha 8, Czechia}
\author[0000-0002-6301-3269]{L\'eo Girardi}
\affiliation{INAF-Osservatorio Astronomico di Padova, Vicolo dell’Osservatorio 5, I-35122 Padova, Italy}
\author[0000-0001-9848-5410]{Bernhard Aringer}
\affiliation{Department of Astrophysics,  University of Vienna, T\"urkenschanzstrasse 17, 1180 Vienna, Austria}
\author[0000-0002-7922-8440]{Alessandro Bressan}
\affiliation{SISSA, via Bonomea 265, I-34136 Trieste, Italy}


\begin{abstract}
In this study we compute the equation of state and Rosseland mean opacity from temperatures of $T\simeq30000$ K down to $T\simeq400$ K, pushing the capabilities of the \texttt{\AE SOPUS} code \citep{aesopus2, Marigo_Aringer_09} into the regime where solid grains can form. The \texttt{GGchem} code \citep{ggchem_18} is used to solve the chemistry for temperatures less than $\simeq 3000$ K. 
Atoms, molecules, and dust grains in thermodynamic equilibrium are all included in the equation of state. To incorporate monochromatic atomic and molecular cross sections, an optimized opacity sampling technique is used. The Mie theory is employed to calculate the opacity of 43 grain species. Tables of Rosseland mean opacities for scaled-solar compositions are provided. Based on our computing resources, opacities for other chemical patterns, as well as various grain sizes, porosity, and shape, can be easily computed upon user request to the corresponding author.
\end{abstract}

\keywords{Stellar atmospheric opacity(1585) --- Astrochemistry(75) --- Silicate grains(1456) --- \\Carbonaceous grains(201) --- Optical constants(2270) }


\section{Introduction} \label{sec:intro}
One of the major theoretical tasks in modern astrophysics is predicting the elemental abundances and opacities in the atmospheres of stars, brown dwarfs, and planets. The interpretation of spectroscopic observations, in particular, necessitates thorough information of these two components.
In recent years, technological advancements have enabled us to investigate the surface/circumstellar chemical composition and spectral features of evolved red giants such as asymptotic giant branch (AGB) stars \citep{Ramstedt_etal_20, Decin_etal_17,  Ramstedt_Olofsson_14}, brown dwarfs \citep{Helling_Casewell_14, Cushing_etal_08, Burrows_etal_98, Marley_97, Allard_etal_97}, and exoplanets \citep{Birkby_etal_17, Fraine_etal_14}, exploiting a wide range of wavelengths.
The James-Webb Space Telescope's infrared equipment is expected to transform our knowledge of these low-temperature objects \citep{Beichman_etal_14}.
In the future,  PLATO \citep{plato_16} will search for habitable, Earth-like planets by applying astroseismology to solar-like stars, for which modeling efforts to derive elemental abundances and opacities will be critical.
The chemistry and opacity of the coolest objects 
must include the contribution of solid species, commonly referred to as dust grains, typically below $T\simeq 1500$ K.

The most used opacity tables that take into account dust grains are those of the Wichita State University group. 
\cite{Alexaner_75} computed opacities down to $T\simeq 700$ K, including a rough estimate of dust grain opacity, while better approximations were later introduced by \cite{Alexander_Ferguson_94}.
In 2005 \citeauthor{Ferguson_etal_05}  made a significant advancement, using the \texttt{PHOENIX} code \citep{Allard_etal_01} to compute the abundances of solid grains in thermal equilibrium within the equation of state (EoS) solution.
Other important efforts include \cite{Semenov_etal_03}'s opacity tables for primary use in protoplanetary disk models, suitable for gas and dust mixtures ranging from $T\simeq 10$ K to $T\simeq 10000$ K, and \cite{Freedman_etal_08} who computed line and mean opacities, without the contribution of dust grains,  for ultracool dwarfs and extrasolar planets in the temperature interval  from $T\simeq 75$ K to $T\simeq 4000$ K.

 In the field of low-temperature gas opacities, \cite{Marigo_Aringer_09} constructed the \texttt{\AE SOPUS} code \citep[][initial version \texttt{\AE SOPUS 1.0}] {Marigo_Aringer_09}, which solves the equation of state for over 800 chemical species (atoms/ions and molecules) and computes the Rosseland mean opacities for arbitrary chemical compositions under the assumption of thermodynamic equilibrium.
 Recently, \citet[][current version \texttt{\AE SOPUS 2.0}]{aesopus2} made a major update to include new thermodynamic data (e.g., partition functions) and to expand molecular absorption to include 80 species,  mostly taken from the \texttt{ExoMol} and \texttt{HITRAN} databases \citep{EXOMOL_2012, HITRAN2022}. The reader should consult the \cite{aesopus2} paper for more information, particularly Tables 1  and 2 for a complete list of all opacity sources considered.

For both versions of \texttt{\AE SOPUS}, we set up a web-interface (\href{http://stev.oapd.inaf.it/aesopus}{http://stev.oapd.inaf.it/aesopus}) that enables users to compute in real time opacity tables based on  their specific needs. The online service provides full control over the chemical abundances of $92$ atomic species, ranging from Hydrogen to Uranium.
Both EoS and opacity calculations are performed over a temperature range of $1500 \lesssim T/{\rm K} \lesssim 30000$, with the chemistry of all components in the gas phase.

In this study we extend the computations to lower temperatures down to  $100$ K, where liquid and solid species appear and dominate both the EoS and the opacity.
The EoS is solved using two codes: \texttt{\AE SOPUS} for temperatures $3000 \lesssim T/{\rm K} \lesssim 30000$ \citep{aesopus2, Marigo_Aringer_09},   and \texttt{GGchem} for temperatures  $100 \lesssim T/{\rm K} \lesssim 3000$ \citep{ggchem_18}. The latter takes into account grain condensation in thermal equilibrium with the gas phase.
Low-temperature opacities are calculated using optical constants for a wide range of materials that condense in the coolest layers of star atmospheres and during star and planet formation. 

New Rosseland mean opacity tables are provided here for scaled-solar abundances. 
They are built as a function of two standard parameters:
\begin{equation}
T\,\,\,\,\,{\rm and}\,\,\,\,\,R=\rho\, T_6^{-3}\,,
\end{equation}
where $T$ is the temperature (in K) and the $R$ parameter contains both the temperature ($T_6=T/(10^6\,{\rm K})$) and the gas mass density $\rho$ (in g\,cm$^{-3}$).
We recall that using the $R$ parameter instead of gas density $\rho$ or pressure $P$ enables the opacity tables to encompass rectangular areas of the $(R,T)$-plane, resulting in an appropriate format for smooth opacity interpolation.

The Rosseland mean opacity tables and optical constants for dust species are available via the repository at \url{http://stev.oapd.inaf.it/aesopus\_2.0/tables}; a copy of these files have also been deposited to Zenodo: \url{https://doi.org/10.5281/zenodo.8221362}.

This paper is organized as follows.
Section~\ref{sec_eos} details  the method for solving the EoS. The opacity of solid grains is treated in Section~\ref{sec_grains} and in Table \ref{tab_grains}. Section~\ref{sec_kross} provides a thorough discussion of the Rosseland mean opacity and summarizes some important updates of molecular line absorption, with a focus on the temperature window where dust grains appear. Opacities for proto-planetary disks, chemical composition effects (for example, alpha-enhanced mixtures) are also investigated, along with a comparison to other available opacity data.
The new opacity tables are introduced in Section~\ref{sec_tables}. A few concluding remarks close the paper in Section~\ref{sec_concl}.
\begin{deluxetable}{ccccccc}[h]
\tabletypesize{\scriptsize}
\tablecolumns{6} 
\tablewidth{0pc} 
\tablecaption{Properties of Condensates} 
\label{tab_grains}
\tablehead{ 
\colhead{Condensate} & \colhead{Name} & \colhead{Group} & \colhead{$\rho_{\rm d} ({\rm g\, cm^{-3}})$} & \colhead{Optical  Constants} & \colhead{$\lambda$ (\micron) range} & \colhead{Analog}
}  
\startdata
am-Al$_2$O$_3$  & Corundum & Ox-/hydroxides & 3.97& 1,39    & $0.2 \leq \lambda \leq 500$  &   \\
MgO           & Periclase    & Ox-/hydroxides & 3.58 & 2      & $0.016 \leq \lambda \leq 625$     &   \\
SiO           & Silicon Monoxide    & Ox-/hydroxides & 2.18 & 3,4      & $0.05 \leq \lambda \leq 100$     &   \\
FeO           & Ferropericlase    & Ox-/hydroxides & 5.99 & 5      & $0.2 \leq \lambda \leq 500$     &   \\
Fe$_2$O$_3$           & Hematite    & Ox-/hydroxides & 5.27& 6      & $0.1 \leq \lambda \leq 1000$     &   \\
Fe$_3$O$_4$           & Magnetite   & Ox-/hydroxides &5.20 & 6      & $0.1 \leq \lambda \leq 1000 $     &   \\
TiO$_2$              & Rutile  & Ox-/hydroxides &4.23 & 7,8,9 & $0.4662 \leq \lambda \leq 36.2$     &   \\
ZrO$_2$             & Baddeleyite  & Ox-/hydroxides & 5.68 & 10 & $4.545 \leq \lambda \leq 25$     &   \\
MgSiO$_3$             & Enstatite & Pyroxenes  & 3.19 & 11     & $0.196 \leq \lambda \leq 9998$   & \\
NaAlSi$_2$O$_6$ & Jadeite & Pyroxenes  & 2.27& 12 & $6.69 \leq \lambda \leq 853$   &   \\
NaAlSi$_3$O$_8$ & Albite & Feldspars  & 2.62& 12 & $6.69 \leq \lambda \leq 853$   &   \\
Ca$_2$Al$_2$SiO$_7$   & Gehlenite   &  Melilites &3.04  & 12   & $6.69 \leq \lambda \leq 853$   &   \\
CaMgSi$_2$O$_6$     & Diopside   &  Pyroxenes& 3.28  &  38         &    $7 \leq \lambda \leq 40$          & \\
CaMgC$_2$O$_6$  & Dolomite & Carbonates & 2.86 & 13 & 
$2.5 \leq \lambda \leq 50$   &   \\
CaSiO$_3$             & Wollasnonite &   Pyroxenoids & 2.91  &   14      &         $0.00282 \leq \lambda \leq 6.198$                             &  \\
Mg$_2$SiO$_4$         & Forsterite   & Olivines & 3.21 & 11     & $0.1957 \leq \lambda \leq 948$   &   \\
Ca$_2$SiO$_4$         & Larnite     &   Nesosilicates & 3.34 &              &                                      &   from Mg$_2$SiO$_4$ \\
Fe$_2$SiO$_4$         & Fayalite    & Olivines & 4.39 & 15     & $0.4\leq \lambda \leq 10000$      &   \\
$\alpha-$SiO$_2$      & Quartz   & Silicates & 2.648 &1,4 & $6.67 \leq \lambda \leq 487.4$     &   \\
MgAl$_2$O$_4$         & Spinel   & Spinels   & 3.58 & 7,16      & $0.35\leq \lambda \leq 10000$         &   \\
CaTiO$_3$             & Perovskite & Perovskite & 3.98 & 8,17  & $0.0356 \leq \lambda \leq 5843$   &   \\   
Na$_2$S  &Sodium Sulfide & Sulfides & 1.86 & 18 
& $0.04 \leq \lambda \leq 200$   &   \\
FeS                   & Troilite    & Sulfides & 4.83 & 19,20     & $0.1 \leq \lambda \leq 487$       &   \\
H$_2$O (ice)          & Water   & Ices & 0.93  & 21      & $0.0443 \leq \lambda \leq 2 \cdot 10^6$       &   \\
NH$_3$ (ice)               & Ammonia  & Ices &0.87  & 22     & $2.5 \leq \lambda \leq 17$     &   \\                    
MgTiO$_3$             & Geikeilite & Ilmenites    & 3.88 &  23      &       $0.25 \leq \lambda \leq 1$             &   \\
NaCl                  & Halite       & Salts & 2.165 & 24     & $0.0477 \leq \lambda \leq 30590$     &   \\
MnS                   & Alabandite  & Rocksalts  & 4.08 & 25,26     &       $0.09 \leq \lambda \leq 190$       &  \\
KCl     &Sylvite & Halides & 1.99 & 27 &  $2 \leq \lambda \leq 1000$                              &  \\
C                     & Graphite     & C-rich matter &2.27  &28  & $0.0001 \leq \lambda \leq 123984$   &   \\
am-C                     & amorphous C     & C-rich matter &2.27  &29  & $19.3 \leq \lambda \leq 50119$   &   \\
SiC                   & Moissanite   & Carbides & 3.21 & 30     & $0.001 \leq \lambda \leq 1000$     &   \\
TiC                   & Titanium Carbide   & Carbides & 4.93 & 31,32     & $0.015 \leq \lambda \leq 207$     &   \\
Ti                    & Titanium       & Metals & 4.14& 33    & $0.667 \leq \lambda \leq 200$     &   \\
Cr                    & Chromium      & Metals &7.19 & 34,35    & $0.04 \leq \lambda \leq 500$     &   \\
Mn                    & Manganese      & Metals &7.43 & 13    & $0.22 \leq \lambda \leq 55.6$     &   \\
Fe                    & Iron      & Metals & 7.87  & 35    & $0.2 \leq \lambda \leq 285.7$     &   \\
Ni                    & Nickel       & Metals&8.91 & 36   & $0.667 \leq \lambda \leq 286 $     &   \\
Cu                    & Copper      & Metals &8.93 & 37      & $0.517 \leq \lambda \leq 55.6$    &   \\
Zn                    & Zinc      & Metals &7.14 & 13      & $0.36 \leq \lambda \leq 55.6$    &   \\
Zr                    & Zirconium      & Metals &6.52 & 13      & $0.22 \leq \lambda \leq 55.6$    &   \\
W                    & Tungsten      & Metals & 19.25& 33     & $0.667 \leq \lambda \leq 200$     &   \\
\enddata
\tablecomments{Column (4) lists $\rho_{\rm d}$, the specific density of the pure substance. In absence of optical constants we use data from analog species in column (7) as in \cite{Ferguson_etal_05}. 
The abbreviation am stands for amorphous. The optical constants for a few species are taken from \cite{Kitzmann_etal_17}.  The entire optical constant dataset can be collected from Zenodo: \url{https://doi.org/10.5281/zenodo.8221362}}
\tablerefs{
(1) \cite{Begemann_etal_97}; 
(2) \cite{Roessler_Huffman_91};
(3) \cite{Wetzel_etal_13};
(4) \cite{Philipp_85};
(5) \cite{Henning_etal_95};
(6) \href{https://www.astro.uni-jena.de/Laboratory/OCDB/index.html}{DOCCD Jena Laboratory}; 
(7) \cite{Zeidler_etal_11};
(8) \cite{Posch_etal_03};
(9) \cite{RIBARSKY1997795};
(10) \cite{Dowling_Randall_77};
(11) \cite{Jager_etal_03};
(12) \cite{Mutschke_etal_98};
(13) \cite{Querry_87};
(14) \cite{Shaker_etal_18};
(15) \cite{Fabian_etal_01};
(16) \cite{Tropf_Thomas_91};
(17) \cite{Ueda_98};
(18) \cite{Khachai_etal_09};
(19) \cite{Henning_Mutschke_97};
(20) \cite{Pollack_etal_94};
(21) \cite{Warren_84};
(22) \cite{Hudson_etal_22};
(23) \cite{Hsiao_etal_11};
(24) \cite{Eldridge_Palik_85};
(25) \cite{Montaner_etal_79};
(26) \cite{Huffman_Wild_67};
(27) \cite{Palik_85};
(28) \cite{Draine_03};
(29) \cite{Jager_etal_98};
(30) \cite{Laor_Draine_93};
(31) \cite{Henning_Mutschke_01};
(32) \cite{Koide_etal_90};
(33) \cite{Ordal_etal_88};
(34) \cite{Rakic_etal_98};
(35) \cite{Lynch_Hunter_91};
(36) \cite{Ordal_etal_87};
(37) \cite{Ordal_etal_85};
(38) \cite{Arnold_etal_2014}
(39) \cite{Koike_etal_95}
}
\end{deluxetable}
\section{Equation of State}\label{sec_eos}
\begin{figure}
    \centering    \includegraphics[width=0.70\textwidth]{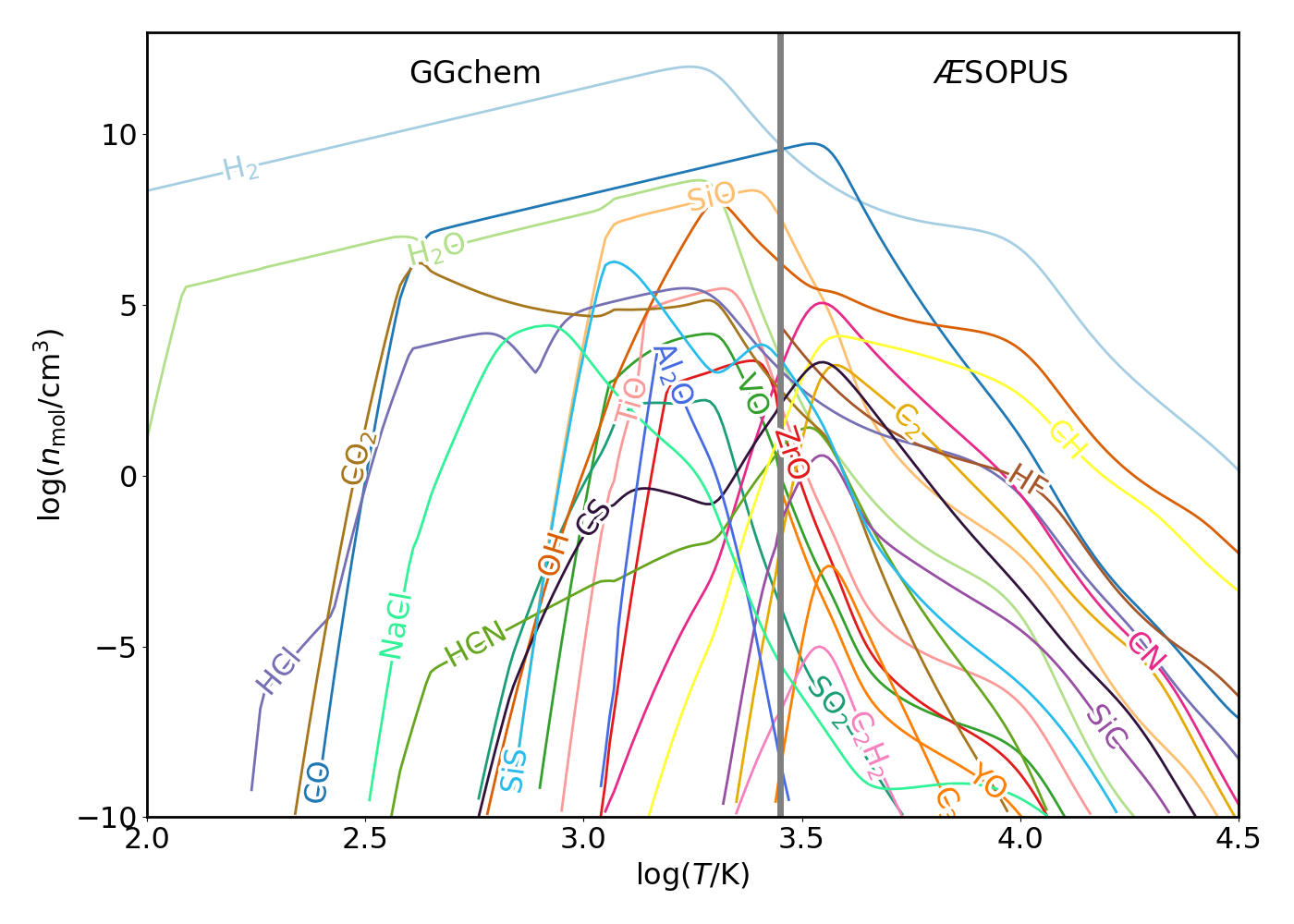}
    \caption{Abundances of a sample of molecules in the gas phase as a function of temperature, and for $\log(R)=-3$.
    The chemical composition assumes
$X = 0.735,\, Z = 0.0165$, with scaled-solar elemental abundances according to \cite{Magg_etal_22}. The gray vertical line defines the transition temperature at $\log(T/{\rm K})=3.45$ between \texttt{\AE SOPUS} and \texttt{GGchem}.}
    \label{fig_molec}
\end{figure}
The software program \texttt{\AE SOPUS} employs the ideal gas assumption to solve the equation of state for over 800 species, including approximately 300 atoms and ions and 500 molecules, in the gas phase under conditions of thermodynamic and instantaneous chemical equilibrium. It encompasses the  temperature range $1500 \lesssim T/{\rm K} \lesssim 30000$.
At low temperatures ($T \lesssim 2000$ K), the problem of determining the equilibrium chemical composition becomes more complex, as it entails also the formation of liquids and solids. In this regime we solve the EoS using the computer code \texttt{GGchem} \citep{ggchem_18}, which computes the abundances of roughly $568$ gas molecules, $55$ liquid species and almost $200$ solid particles.
We set a transition temperature of $\simeq 3000$ K, below which we switch from \texttt{\AE SOPUS} to \texttt{GGchem}.

Figure \ref{fig_molec} depicts the concentrations of a few molecules in the gas phase as a function of temperature. As we can see, the match between \texttt{\AE SOPUS} and \texttt{GGchem} is  smooth and without discontinuities across the transition temperature. 
To better assess the differences in molecular abundances between \texttt{\AE SOPUS} and \texttt{GGchem}, we performed two independent runs of the codes in the temperature interval $3.3 \leq \log(T/{\rm K}) \leq  3.7$.
 We find that deviations in the predicted molecular concentrations typically range from a few 0.001 dex to a few 0.01 dex.
This gives us confidence in the physical consistency of the two codes.

\begin{figure}[H]
    \centering    \includegraphics[width=0.60\textwidth]{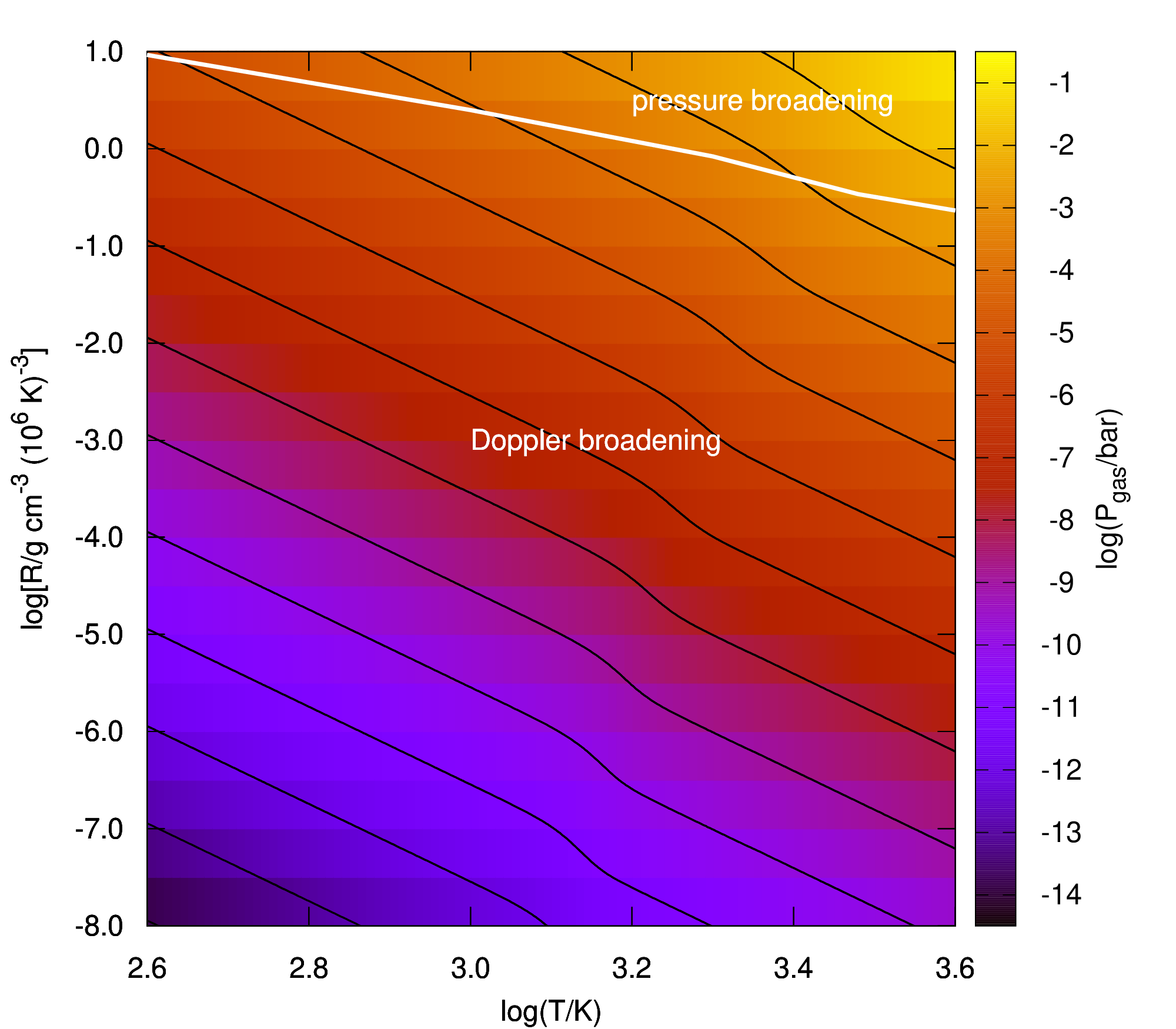}
    \caption{Map of gas pressure (bar) for temperatures $380 \lesssim T/{\rm K}\lesssim 4000$, where molecules and dust grains dominate the opacity. Contour levels (black lines) are distributed every 1 dex in $\log(P)$. The thick white line marks the boundary above which pressure broadening  of molecular spectral  absorption starts to affect the line wings \citep{Hedges_Madhusudhan_16}. Below that line, thermal Doppler effect can be safely assumed.}
    \label{fig_pgas}
\end{figure}

Additionally, it is important to examine the gas pressure that our opacity tables cover. Figure~\ref{fig_pgas} provides an example. The  chemical composition is  scaled-solar according to \cite{Magg_etal_22} with total metallicity $Z=0.02$, and hydrogen abundance $X=0.7$.  \citet[][see their figure 2]{Hedges_Madhusudhan_16} carried out a thorough investigation to identify the relative dominance of pressure and Doppler broadening mechanisms in the pressure-temperature space. 
Following that analysis we indicate an approximate limit above which molecular lines start to be broadened by pressure, resulting in a Lorentzian and/or Voigt profile \citep{Burrows_etal_98}. We observe that only the top right corner of the opacity tables are affected by this phenomenon, whereas thermal Doppler is generally the main broadening mechanism. In terms of atomic opacities, we recall that we use  monochromatic cross sections from the \texttt{Opacity Project}  \citep{Seaton_etal_94}, where line broadening is caused by pressure, radiation damping, and thermal Doppler effects.

\begin{figure}[H]
    \centering    \includegraphics[width=0.70\textwidth]{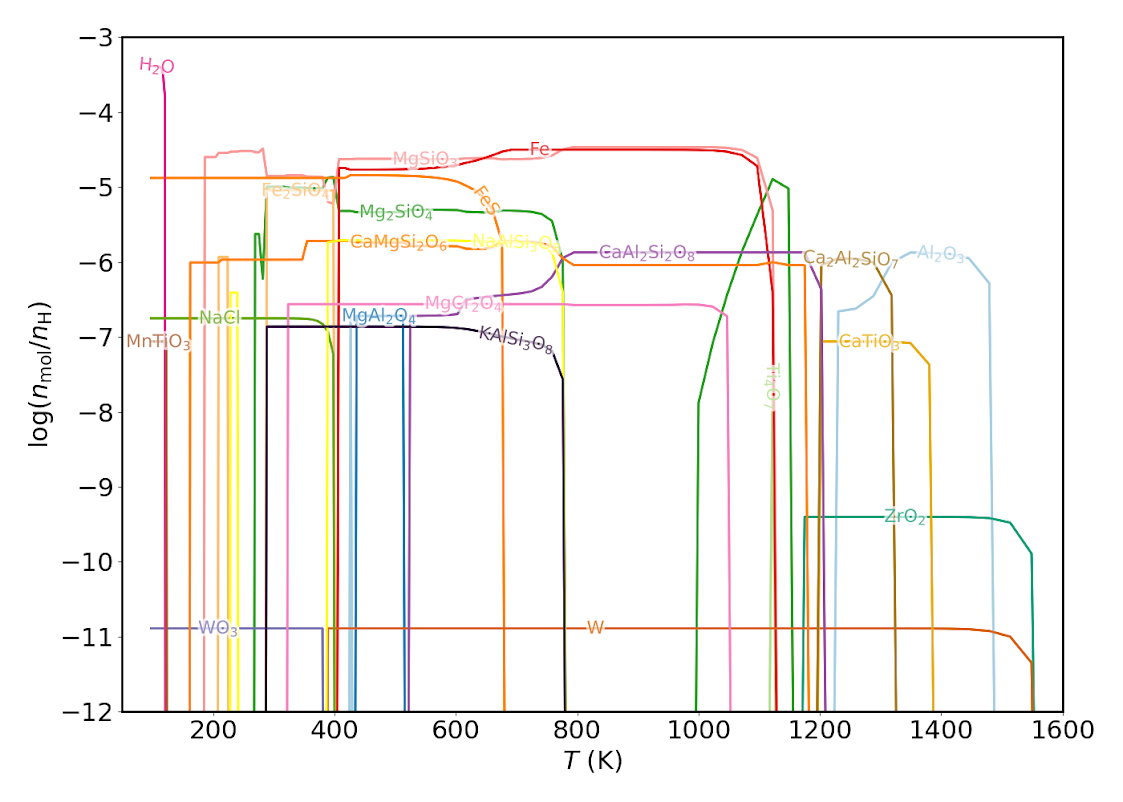}
    \caption{Onset of condensation as a function of temperature at $\log(R)=-3$, for solar abundances in phase equilibrium, computed with \texttt{GGchem}. The chemical composition is the same as in Fig.~\ref{fig_molec}.
    The plot depicts the abundances of  several condensed species with respect to hydrogen nuclei.}
    \label{fig_cond}
\end{figure}

Figure~\ref{fig_cond}  illustrates a sample of condensed species  for a scaled-solar composition with $\log(R)=-3$. Moving down to lower temperatures, the gas density ranges from $2.1\cdot 10^{-3}$ to $9.4\cdot 10^{-16}\, {\rm g\,cm^{-3}}$, while the gas pressure ranges from $3.8$ to $3.2\cdot 10^{-6}\, {\rm dyne\,cm^{-2}}$.
The general trends at a gas pressure of 1 bar are extensively discussed by \cite{ggchem_18}.
In this case, density and pressure are much lower, and condensation begins at lower temperatures. The first stable condensates to form  are crystalline Tungsten (W[s]) and Baddeleyite (ZrO$_2$) at  $T\simeq 1550$ K, Corundum (Al$_2$O$_3$) at $T\simeq1484$ K, Perovskite (CaTiO$_3$) at $T\simeq1386$ K, and Gehlenite (Ca$_2$Al$_2$SiO$_7$) at $T\simeq1326$ K.
Below 1222 K, Al$_2$O$_3$ disappears and is replaced by the Ca-silicates group (CaMgSi$_2$O$_6$, Ca$_2$Al$_2$SiO$_7$, CaAl$_2$Si$_2$O$_8$, and Ca$_2$MgSi$_2$O$_7$), the titanates (CaTiO$_3$), and 
Forsterite (Mg$_2$SiO$_4$) At and below temperatures of about 1125 K, silicates, iron condensates (Fe, FeS) and  other species appear (Al$_6$Si$_2$O$_{13}$, Al$_2$SiO$_5$, 
KAlSi$_3$O$_8$ and NaAlSi$_3$O$_8$, Fe, MgCr$_2$O$_4$,
 MgAl$_2$O$_4$) and contribute the most to the grain abundance down to $T\simeq 160$ K. We note that at  $T\simeq 400$ K Halite (NaCl) start to condense. At lower temperatures the element condensation is completed by one major species: water ice (H$_2$O) at $T\simeq 120$ K. At these low densities ammonia ice (NH$_3$) does not condense in appreciable amounts.

\section{Opacity of Solid Grains} 
\label{sec_grains}
Computing the opacity caused by solid dust grains requires knowledge of the abundances of the different species as a function of temperature and density, as well as the absorption and scattering properties of each individual dust grain. The monochromatic cross section per unit mass $\mathrm{( cm^2\,g^{-1})}$ of a given solid species is calculated with

\begin {equation}
\kappa_\lambda^{ \rm grain} ~=~\frac{\int_{a_{\rm min}}^{a_{\rm max}} n(a)~Q_{\rm ext}~(a,\lambda)~\pi a^2~da}{\rho}\, .
\label{eq_kdust}
\end {equation}
Here we assume that the grains are spherical particles with radius $a$ and follow a power-law size distribution 
\begin{equation}
n(a) da \propto a^{-\alpha}\,,
\label{sizedist}
\end{equation}
 with $\alpha=3.5$ discovered  in the interstellar medium by \cite[][MRN]{Mathis_etal_77}. The lower and upper limits of the size distribution are set to $a_{\rm min}=0.00625\,\micron$ and $a_{\rm max}=0.24\, \micron$, respectively, as determined by MRN for interstellar grains and also adopted by \cite{Ferguson_etal_05}.
While  an interstellar size distribution has been used as a standard in several studies,
we recognize that it does not always apply in other physical situations. 
The total surface area of a dust species, which is proportional to its opacity, is highly dependent on $a_{\rm min}$. 
For instance, more appropriate parameters for modeling dust opacities in protoplanetary disks would be $a_{\rm min}=0.1\,\micron$, $a_{\rm max}=3\, {\rm mm}$ \citep{Woitke_etal_19}. A simple application is discussed in Section~\ref{ssec_proplyds}.
We also plan to vary the size range in follow-up works tailored to specific applications.

The dimensionless quantity $Q_{\rm ext}$ is the total extinction efficiency which includes the absorption efficiency, $Q_{\rm abs}$, and the scattering efficiency, $Q_{\rm scat}$.
The grain extinction efficiencies are calculated applying Mie theory, the key component of which is the complex refractive index (or optical constants). The sources of optical constants, as well as other dust grain properties, are listed in Table~\ref{tab_grains}.
\begin{figure}[h!]
    \centering  
\includegraphics[width=0.48\textwidth]{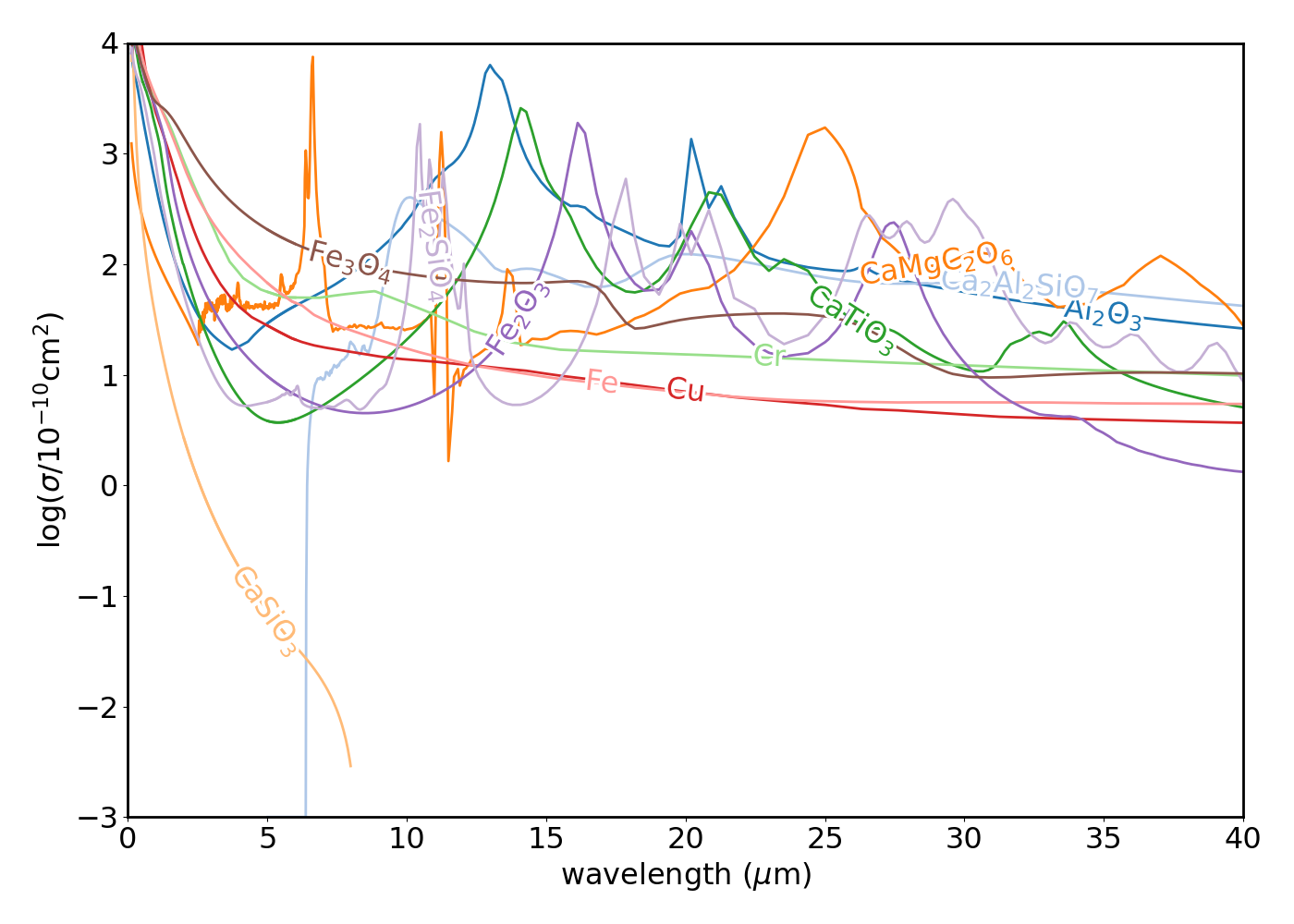}
\includegraphics[width=0.48\textwidth]{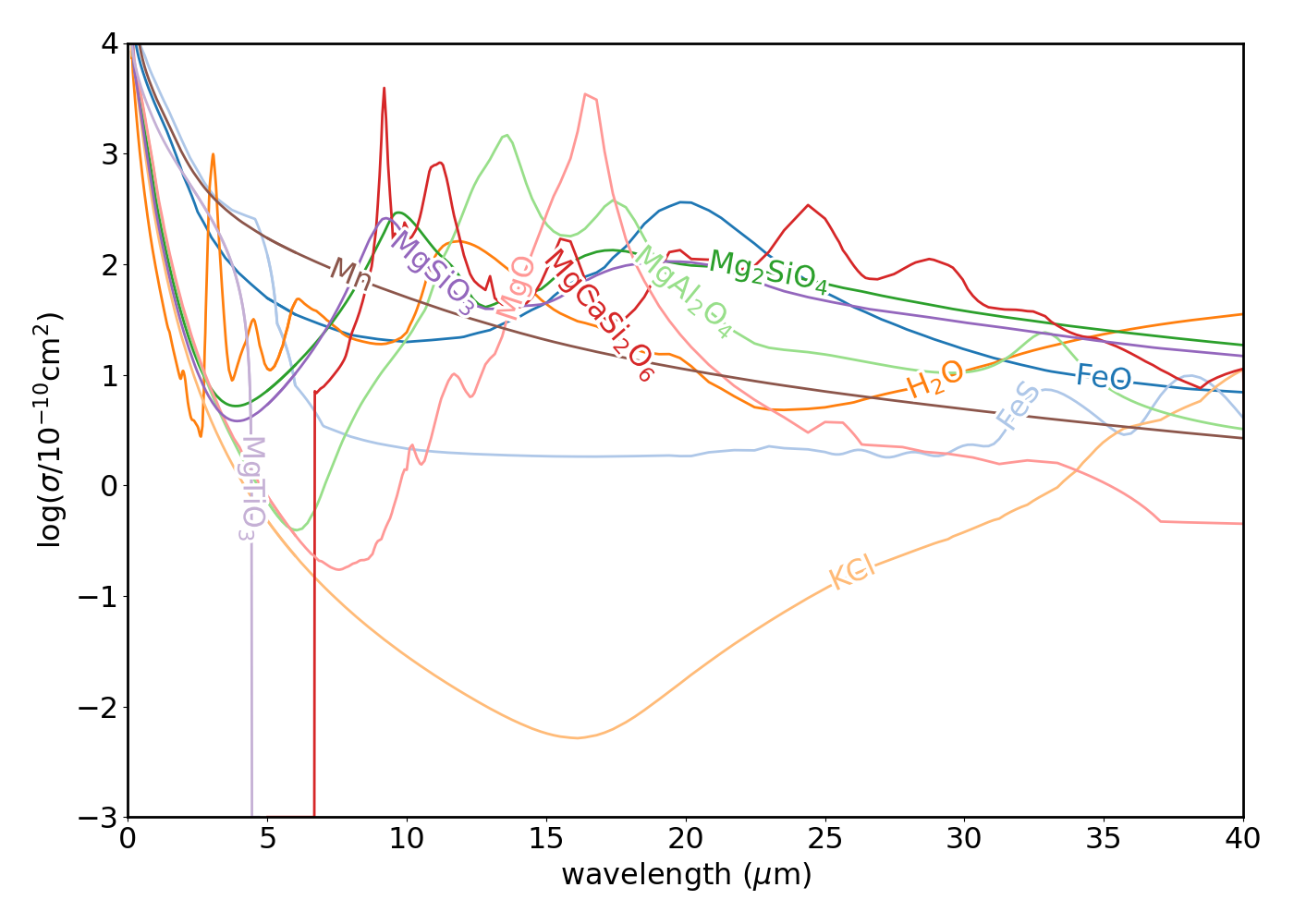}
\includegraphics[width=0.48\textwidth]{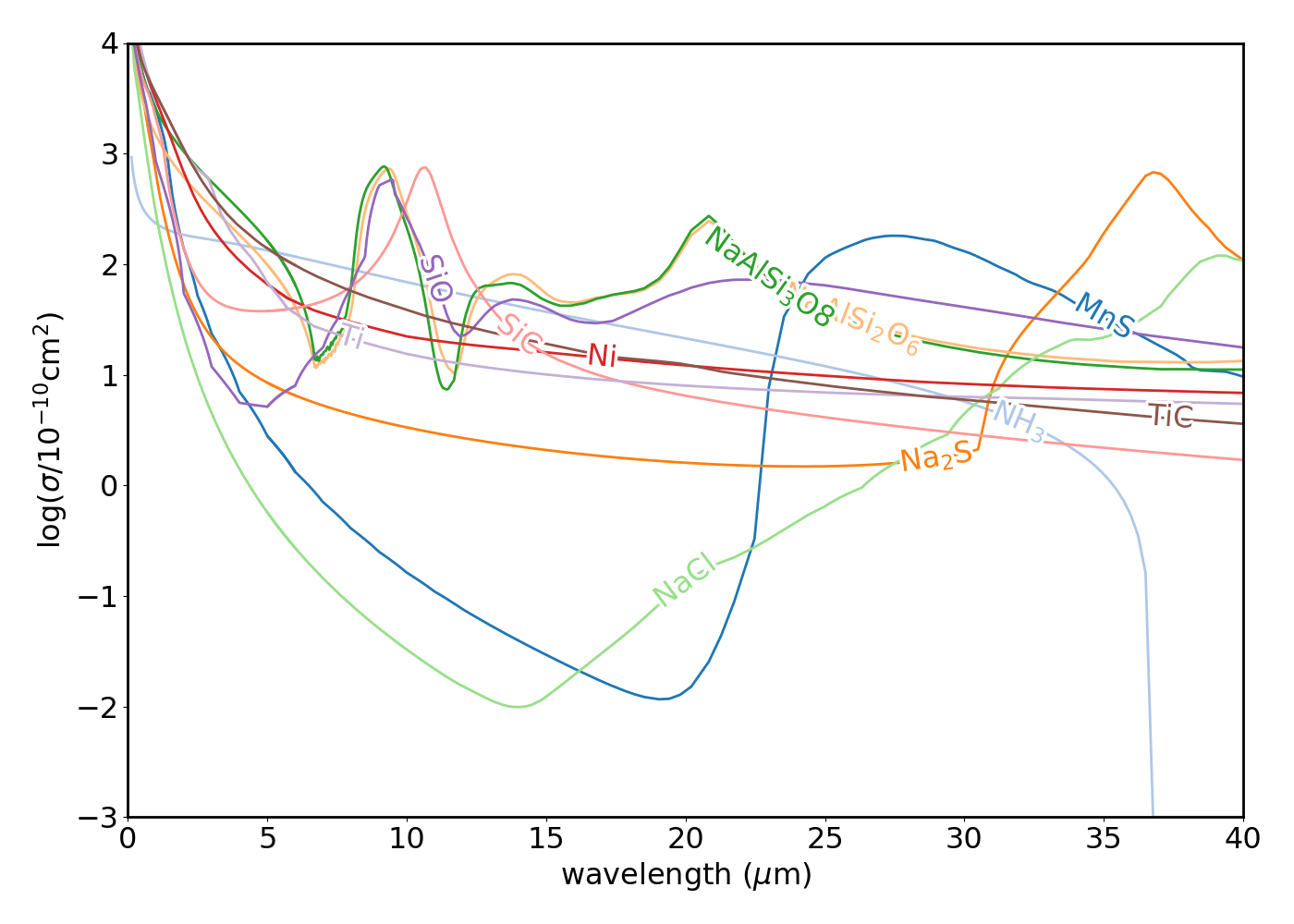}
\includegraphics[width=0.48\textwidth]{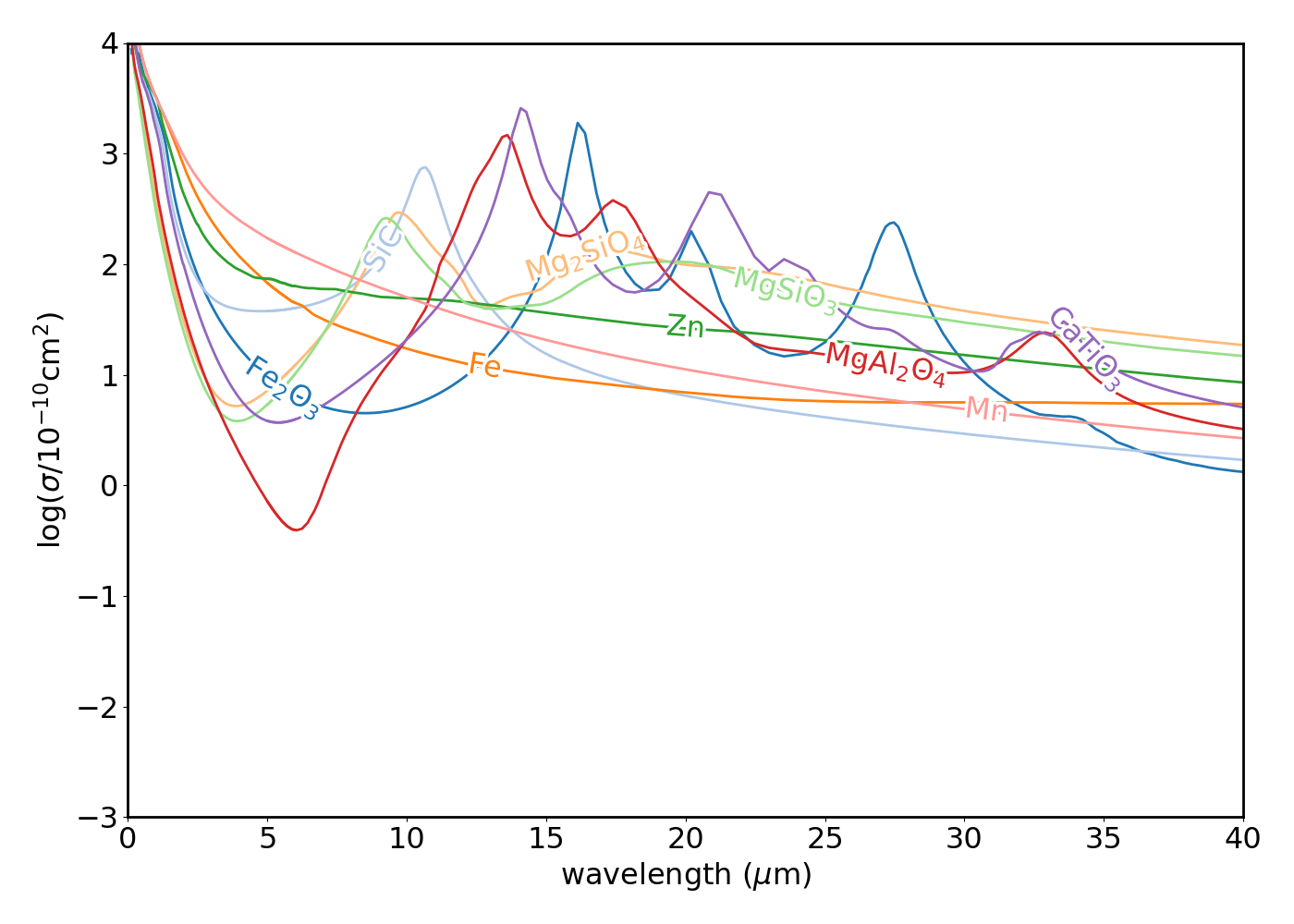}
    \caption{The average size-integrated extinction cross sections of various dust species  made of a single substance. The Mie formalism is used, with diameters ranging from $0.00625\,\micron$ to $0.24\,\micron$ and assuming a power-law grain size distribution (see Equation~\ref{sizedist}).}
    \label{fig_sigma}
\end{figure}

To compute $\kappa_\lambda^{\rm grain}$ of Equation~(\ref{eq_kdust}), we use the \texttt{DIANA} fortran package\footnote{\url{https://diana.iwf.oeaw.ac.at/}} \citep{Woitke_etal_16}.
The \texttt{DIANA} is a versatile code with multiple entry options.
One can specify the minimum and maximum sizes, $a_{\rm min}$ and $a_{\rm max}$, the size distribution's power-law $\alpha$, the porosity $P$ which defines the volume fraction of vacuum, and the distribution of hollow spheres (DHS) with a maximum hollow volume ratio $V^{\rm max}_{\rm hollow}$. As a starting choice we consider  solid homogeneous spheres made of a single substance ($P=0$, $V^{\rm max}_{\rm hollow}=0$). Opacities with varying grain sizes, shape and porosity will also be analyzed in a follow-up work and can be incorporated upon user's request.

The monochromatic extinction profiles of the dust species included in \texttt{\AE SOPUS 2.0} are depicted in Figure~\ref{fig_sigma}.
Monoatomic grains such as Fe, Cu, Cr, Ni,Zn, Zr  contribute scattering mainly at optical wavelengths, whereas Corundum, Spinel, Perovskite, Hematite, Magnetite, Dolomite, and Gehlenite grains exhibit strong absorption peaks at infrared wavelengths.

\section{Rosseland Mean Opacity} \label{sec_kross}

\paragraph{Updates in \AE SOPUS opacity} \texttt{\AE SOPUS} Rosseland mean opacities have previously been computed in the temperature range $1600 \lesssim T/{\rm K} \lesssim 30\,000$. When the temperature is reduced to 400 K, the monochromatic opacities of the molecules must be expanded to cover the appropriate interval. The absorption data for 80 molecules in the gas phase are extended from $T=100$ K to $T\simeq 30\,000$ K. The adopted line lists are the \texttt{EXOMOL} database's recommended ones \citep{EXOMOL_2012}, with a few additions from  \texttt{HITRAN} \citep{HITRAN2022}. Table 2 of \cite{aesopus2} contains the complete references, but for one exception concerning ZrO opacity.
We use the most recent \texttt{ZorrO} linelist \citep{Perri_etal_23}, which has a temperature range of up to 10\,000 K.

Molecular line profiles are treated with a thermal Doppler broadening plus  a micro-turbulence velocity according to the following equations:
\begin{equation}
\phi(\nu)=\frac{1}{\Delta_{\nu}\sqrt{\pi}}\,e^{-\left(\frac{\nu-\nu_0}{\Delta_{\nu}}\right)^2}\,,
\end{equation}
where $\nu_0$ is the line center position in frequency, and $\Delta_{\nu}$ is the line width, obtained with
\begin{equation}
\Delta_{\nu}=\frac{\nu_{0}}{c}\sqrt{\frac{2 k_{\rm B}T}{m}+\xi^2}.
\label{eq_dopplermicro}
\end{equation}
In this Equation $c$ stands for the speed of light, $k_{\rm B}$ for the Boltzmann constant, $m$ for the molecule's mass and $\xi$ for  the micro-turbolent velocity, which is set to $2.5$~km/s \citep[see][for more details]{aesopus2, Marigo_Aringer_09}.
In the range of pressures covered by our computations this should not
significantly alter the Rosseland mean opacity (except for $0 \le \log(R) \le 1$; see Figure~\ref{fig_pgas}), given
that the many different opacity sources overlap in ways that
reduce the impact of ignoring the far wings of molecular transitions.
Neglecting the line-extended wings  could be severely incorrect in the case of planetary, brown dwarf and very low mass star atmospheres with little to no ionization and H primarily appearing in molecular form, H$_2$ \citep{Burrows_etal_01}.

The method for calculating the Rosseland mean opacity is fully described in \cite{aesopus2} and  \cite{Marigo_Aringer_09}.  To recap, for any selected $(\rho, T)$ pair, we first compute the total monochromatic opacity cross section per unit mass (in cm$^2$ g$^{-1}$), by including all the contributions from true absorption and scattering. 
 The difference between this work and previous ones is that we now add $\kappa_\lambda^{ \rm grain}$ of Equation~(\ref{eq_kdust}). Second, we compute the Rosseland mean opacity, $\kappa_{\rm R}$ (in cm$^2$ g$^{-1}$), by integrating over frequency:
\begin{equation}
  \frac{1}{\kappa_{\rm R}(\rho,T)} = \displaystyle\frac{\displaystyle\int_0^\infty \displaystyle\frac{1}{\kappa(\nu)}
\frac{\partial B_\nu}{\partial T} d\nu}{\displaystyle\int_0^\infty \displaystyle\frac{\partial B_\nu}{\partial T} d\nu}\,,
\label{eq_rosseland}
\end{equation}
which is a harmonic weighted average with weights equal to the temperature derivatives of the Planck distribution, 
\text{$\frac{\partial B_\nu}{\partial T}$}.
To set up the frequency grid we use the \citet{Helling_Jorgensen_98} algorithm which optimizes the frequency distribution in the opacity sampling technique (see the thorough discussion in \citealt{aesopus2}).

\begin{figure}[h!]
    \centering
    \includegraphics[width=0.60\textwidth]{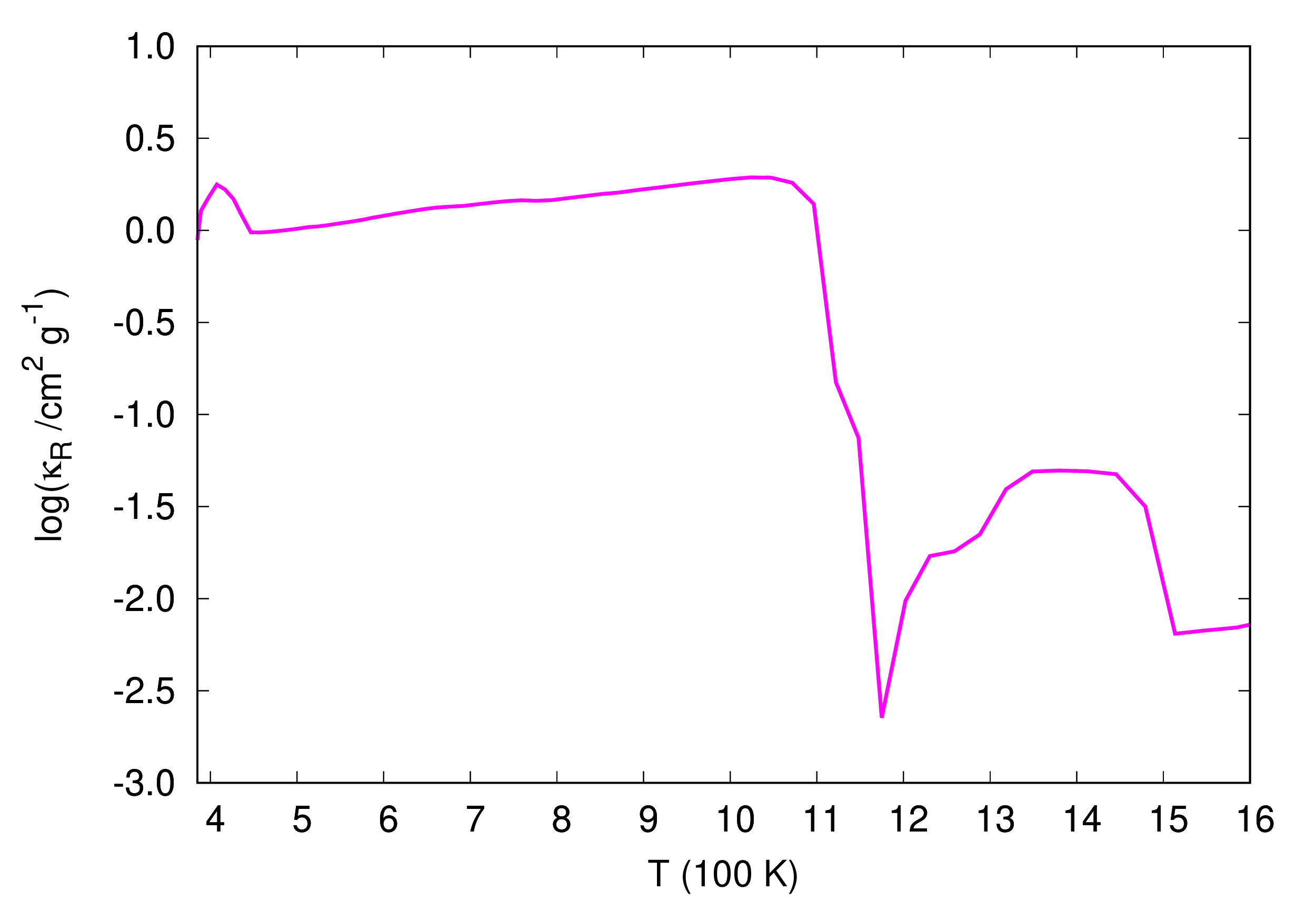}
    \includegraphics[width=0.60\textwidth]{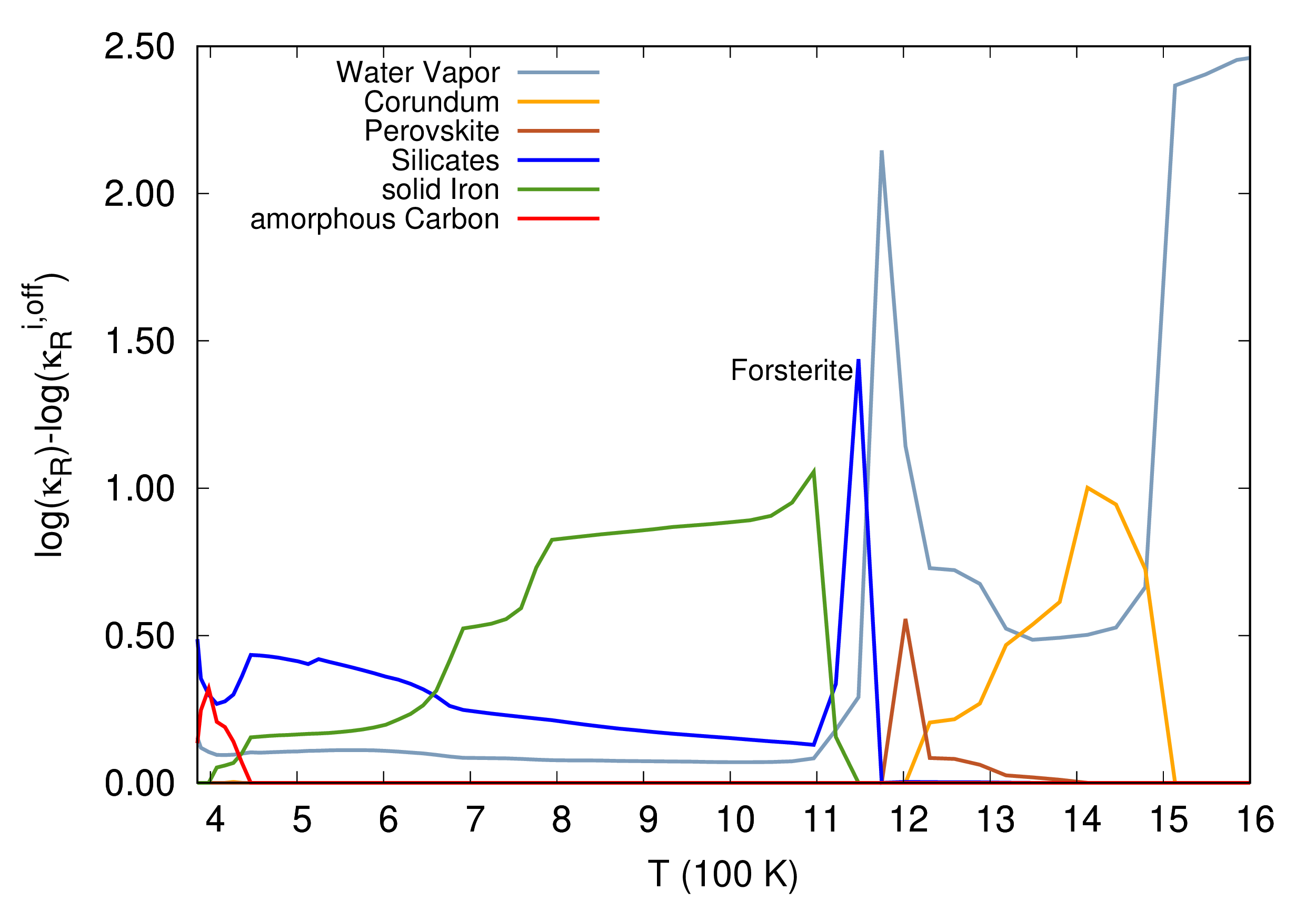}
    \caption{Properties of the Rosseland mean opacity in the low-temperature regime where solid species dominate.  The chemical composition is defined by $X = 0.735,\, Z = 0.0165$, with scaled-solar elemental abundances following \cite{Magg_etal_22}. We take $\log(R)=-3$.
     Top panel: Rosseland mean opacity computed with \texttt{\AE SOPUS~2.0}.
     Bottom panel: contributions to the total Rosseland mean opacity of major solid species.  Each curve corresponds to $\log(\kR)-\log(\kR^{i, {\rm off}})$, where $\kR$ is the full opacity including all opacity sources considered here, and $\kR^{i, {\rm off}}$ is the reduced opacity computed by omitting the specific absorbing species.}
    \label{fig_kgrains}
\end{figure}
 Top panel of Figure~\ref{fig_kgrains}  zooms in the temperature window where solid condensates dominate the Rosseland mean opacity.  Similarly to \cite{aesopus2} and \cite{ Marigo_Aringer_09}, we show  in the bottom panel of Figure~\ref{fig_kgrains}  the quantity $\log(\kR)-\log(\kR^{i, {\rm off}})$ to highlight the temperature windows where the various opacity sources make a significant contribution. Here $\kR$ is the total Rosseland mean opacity including all opacity sources considered here, and $\kR^{i, {\rm off}}$ is the reduced opacity obtained by ignoring the species $i$ the role of which we intend  to investigate.

We notice that \kR\ has abrupt steep rises and drops, which correspond to sudden phase transitions/disappearances of various solid species. An opacity bump appears in the temperature range  
 $1500 \gtrsim T/{\rm K} \gtrsim 1200$, which is caused primarily by the formation of Corundum. At these temperatures, molecular band absorption by water continues to contribute, extending down to $\simeq 400$K. Perovskite
 is responsible for a small spike at $T \approx 1200$ K.
 For temperatures in the interval  $1200 \gtrsim T/{\rm K} \gtrsim 400$ silicates and solid Iron contribute most to \kR.
 At $T\simeq 1130$ K, Forsterite makes a significant contribution. Finally, it is worth noting that amorphous carbon exhibits a moderate  but discernible  opacity bump in \kR\ around $T\simeq 400$ K, in a mixture with solar composition.
 
\begin{figure}[h!]
    \centering
    \includegraphics[width=0.48\textwidth]{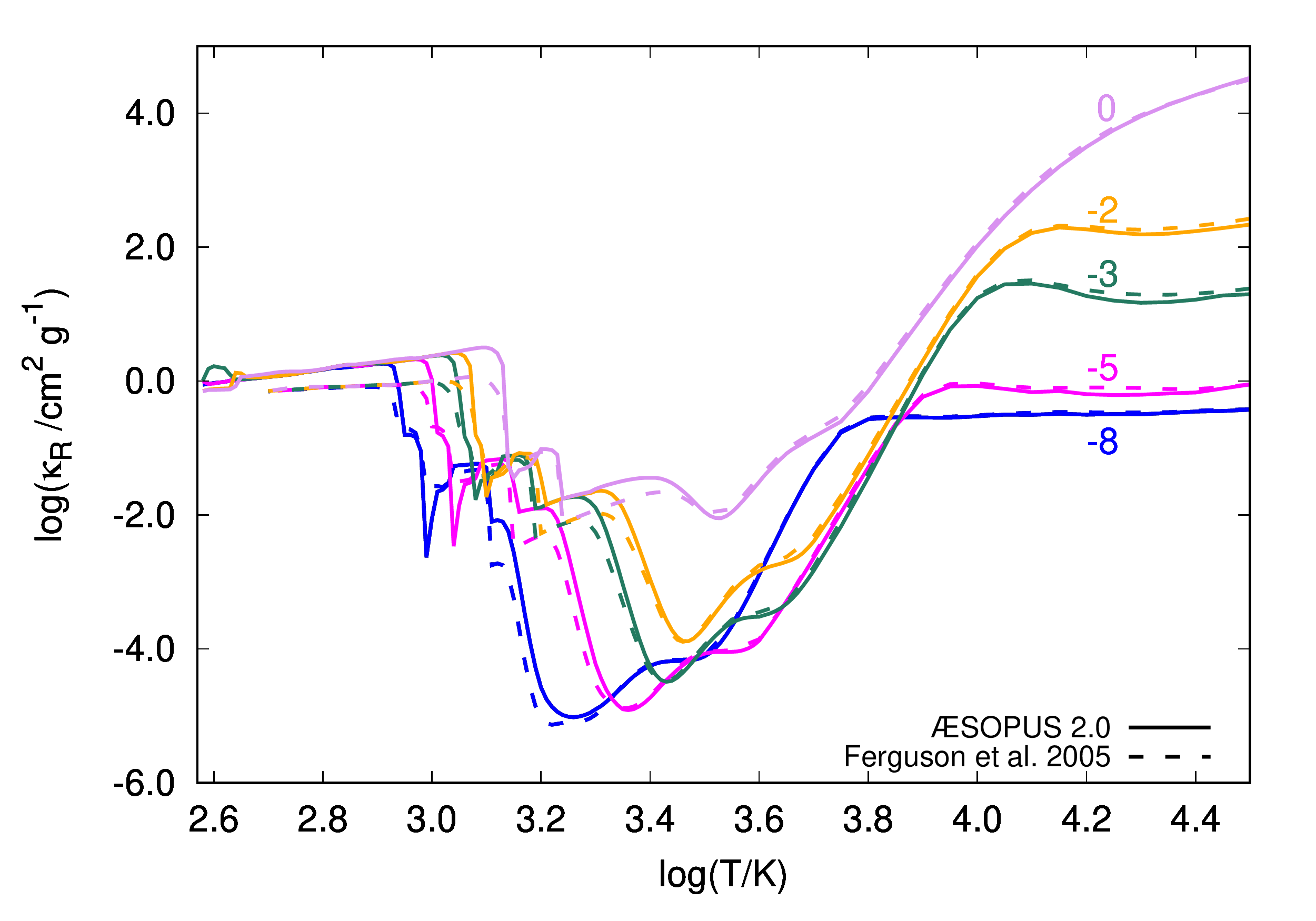}
    \includegraphics[width=0.48\textwidth]{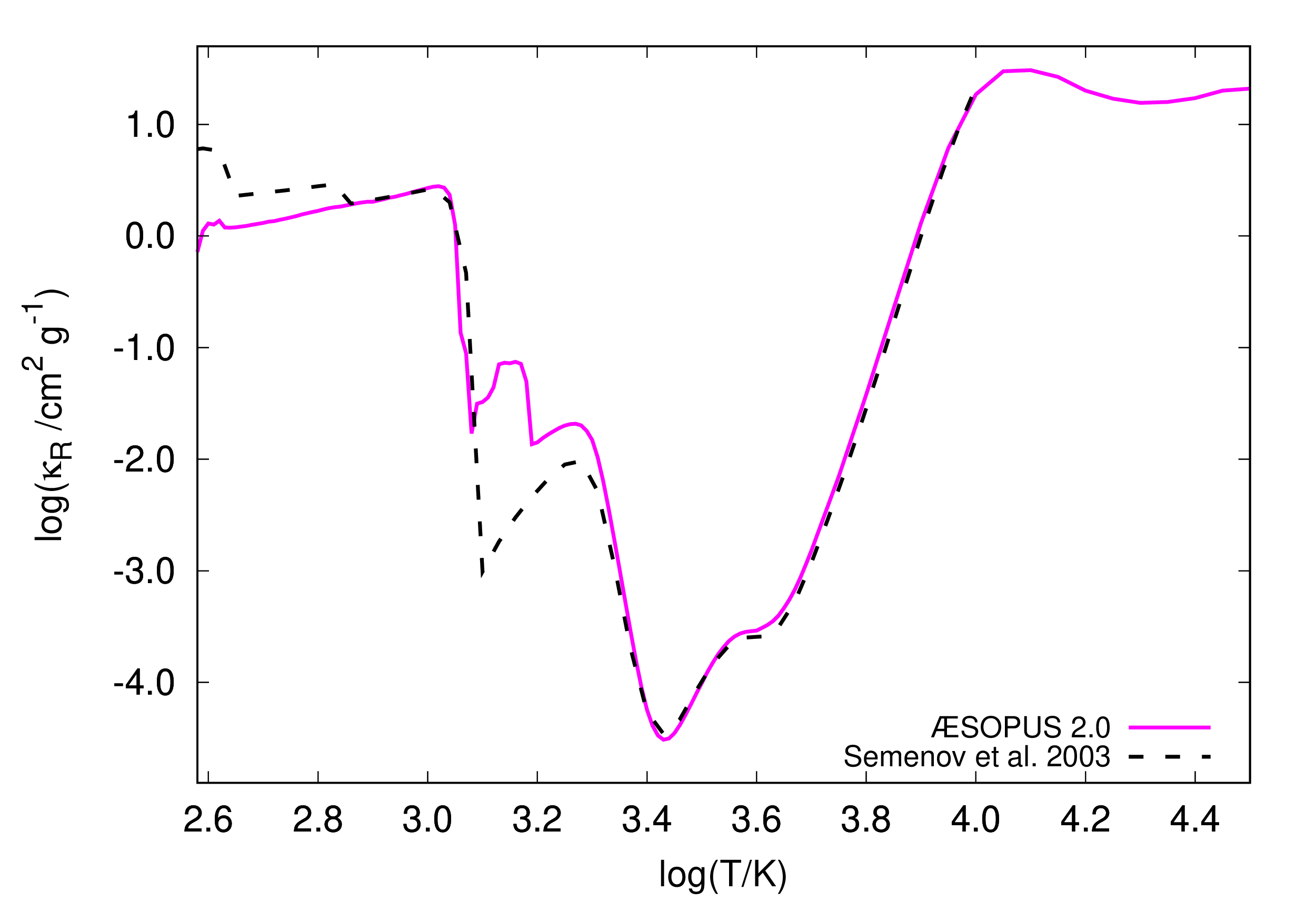}
    \caption{Rosseland mean opacity comparison, between this work and other studies.
    Left panel: comparison with \cite{Ferguson_etal_05}. The chemical composition assumes $X = 0.7,\, Z = 0.02$, with scaled-solar elemental abundances according to \cite{GS_98}, for a few values of the $\log(R)$ parameter which are labeled near the corresponding curves.
    Right panel: comparison with \cite{Semenov_etal_03}. The chemical mixture is defined by $X = 0.732,\, Z = 0.0194$, with scaled-solar elemental abundances according to \cite{AG_1989}. 
    We take $\log(R)=-3$, which corresponds to a density range $10^{-13.2} \le \rho \le 10^{-7.5}$, moving from $\log(T)=2.6$ to $\log(T)=4.5$.}
    \label{fig_ferguson_semenov}
\end{figure}
\subsection{Comparison with other authors}
\label{ssec_others}
Figure~\ref{fig_ferguson_semenov} (left panel) shows a comparison of the results of this work and those of \cite{Ferguson_etal_05}, for a few values of the $R$ parameter. There is a high degree of agreement down to $T\simeq 1600$ K. Below, in the regime of solid grains, differences begin to appear, involving primarily the opacity contributions of Corundum, solid Iron, and Silicates.
These differences could be attributed to different EoS solutions as well as differences in the complex refractive index of the various species.
We note that our computations predict a higher \kR\ produced by silicates. In the silicate regime ($T < 1500$ K), the opacity is slightly affected by the $R$ parameter, whereas noticeable differences appear for the Corundum bump (which condenses at lower temperatures as $R$ decreases) and become more pronounced at higher temperatures.
Another distinction is that we compute opacity down to $T\simeq 400$ K, whereas \cite{Ferguson_etal_05} stop at $T\simeq 500$ K.

The right panel of figure~\ref{fig_ferguson_semenov} compares the results of this work with those of \cite{Semenov_etal_03}. For this latter work, we use the open-source code\footnote{\url{https://www2.mpia-hd.mpg.de/~semenov/Opacities/opacities.html}}  to compute \kR, and we use the assumptions of iron-poor silicates and dust grains considered as homogeneous spheres.
Larger differences emerge in this case. First of all, we predict a higher water opacity bump at $T\simeq\,2000$ K, most likely due to the use of different line lists.  Moreover, the absence of the opacity bump at $T\simeq 1500$ K  in \cite{Semenov_etal_03} is explained by the exclusion of high-temperature condensates, such as Al$_2$O$_3$. 
The discrepancy between the two opacity predictions becomes more pronounced at $T < 1000$ K.
Several factors should be connected to the cause. While \texttt{GGchem} computes grain abundances for each combination of $(T,\rho)$ in thermal equilibrium with the gas phase, \cite{Semenov_etal_03} assumes some fixed abundances characteristic of proto-planetary disks \citep[see also][]{Pollack_etal_94}. Moreover, different approaches are used to define the condensation and vaporization temperatures. Furthermore, the size distribution functions of the grains in the two studies cover very different ranges. In this work we adopt $a_{\rm min}=0.00625\,\micron$ and $a_{\rm max}=0.24\, \micron$, whereas \cite{Semenov_etal_03} take much larger grains, with $a_{\rm min}=0.5\,\micron$ and $a_{\rm max}=5\, \micron$. Finally, different optical constants could be also a contributing factor.

\subsection{Changing dust parameters: a simple application for proto-planetary disks}
\label{ssec_proplyds}
The opacity tables  of this work are computed using a predefined set of physical assumptions for grain physics, based on MRN study for interstellar grains, that may or may not be appropriate for specific applications. The \texttt{DIANA} package \citep{Woitke_etal_16} allows us to adjust our grain assumptions regarding  size and its statistical distribution, porosity, and shape. 
Here we present a simple opacity test customized for the case of proto-planetary disks (proplyds). 
Following recent theoretical works on proplyds \citep{Woitke_etal_19, Woitke_etal_16}  that reproduce continuum and line observations, we extract calibrated dust parameters. Specifically we set $a_{\rm min}=0.1\,\micron$, $a_{\rm max}=3\, {\rm mm}$, porosity $P=0.25$, and $V^{\rm max}_{\rm hollow}=0.8$, and assume power-law size distribution with $\alpha=4$ (for a definition of the quantities, refer to Section~\ref{sec_grains}). For this basic test  we do not take dust settling into account.
 \begin{figure}[h!]
    \centering    \includegraphics[width=0.48\textwidth]{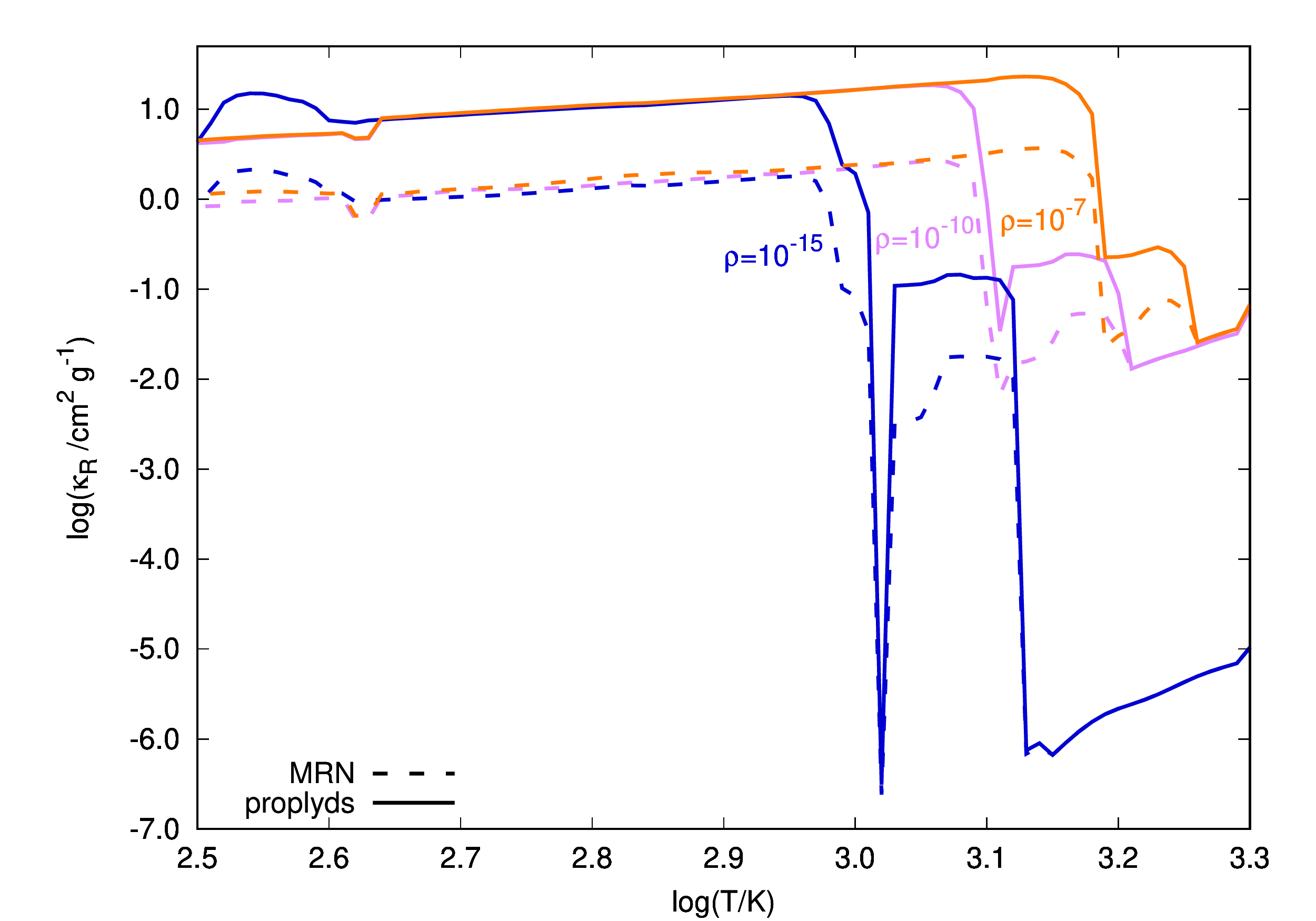}
\includegraphics[width=0.48\textwidth]{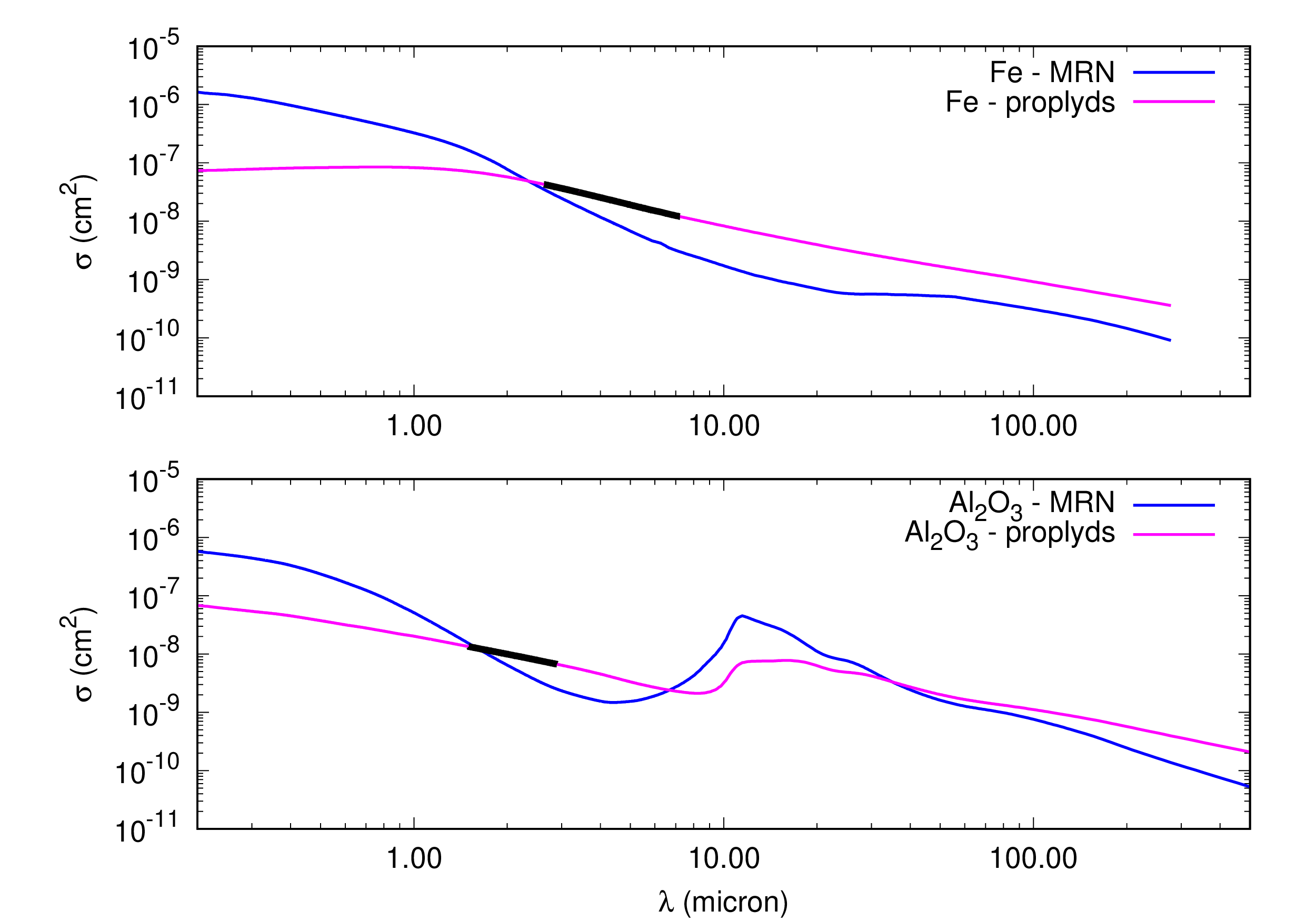}
    \caption{Left panel: Rosseland mean opacities in the dust regime computed with two different sets of assumptions: our standard choice based on MRN (dashed lines), and modified dust parameters appropriate for proplyds (solid lines). 
    Each line represents a constant gas mass density  within the \cite{Semenov_etal_03} specified range.
    The chemical composition is scaled-solar according to \cite{Magg_etal_22}, with metallicity $Z=0.02$ and hydrogen abundance $X=0.7$.
    Right panel: size-integrated extinction cross sections for Iron and Corundum for two dust prescriptions examined in this work. According to  Wien's displacement law, thick black lines correspond to the maximum wavelengths at the typical temperatures where the two dust species contribute most to the opacity. See text for more details.}
    \label{fig_dustchange}
\end{figure}
Figure~\ref{fig_dustchange} (left panel) compares dust opacities computed according to MRN prescriptions for interstellar grains, and those suitable for proplyds. We note that Rosseland opacities for proplyds are  much higher than those computed with MRN prescriptions, up to a factor of 5.6 in the temperature regime, below 1000 K, where silicates and iron dominate the opacity.
We caution that the differences could be lessened by including dust settling.
To explain the reason, in the right panel of Figure~\ref{fig_dustchange} we compare the size integrated cross sections as a function of  wavelength. The curves exhibit  variable trends, and cross sections for proplyds can be higher or lower than MRN cross sections at different wavelength ranges. However, using the Wien's displacement law we can roughly estimate the peak wavelengths of the spectrum (black thick lines) at  the typical temperatures where Corundum ($1000 \lesssim T/{\rm K} \lesssim 2000$) and Iron ($400 \lesssim T/{\rm K} \lesssim 1000$) contribute significantly to the opacity. This clarifies that proplyds have  higher Rosseland mean opacities than MRN, because their cross sections are larger under these circumstances. Similar patterns are seen for silicates, as well as for the amorphous carbon opacity bump, particularly evident at $300 \lesssim T/{\rm K} \lesssim 400$ for $ \rho=10^{-15}$ g cm$^{-3}$.

\subsection{Composition Effects}
\begin{figure}[h!]
    \centering    \includegraphics[width=0.48\textwidth]{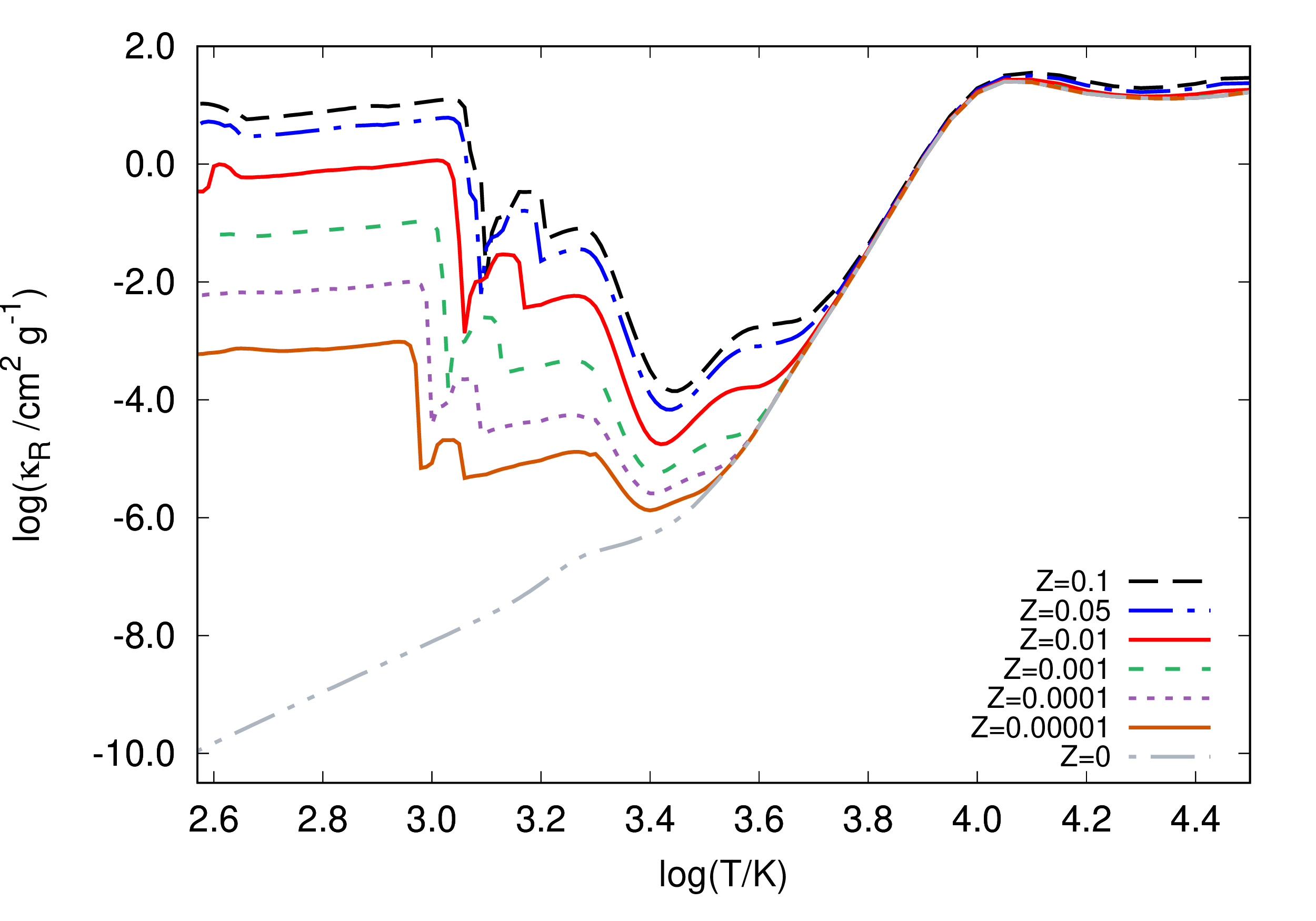}    \includegraphics[width=0.48\textwidth]{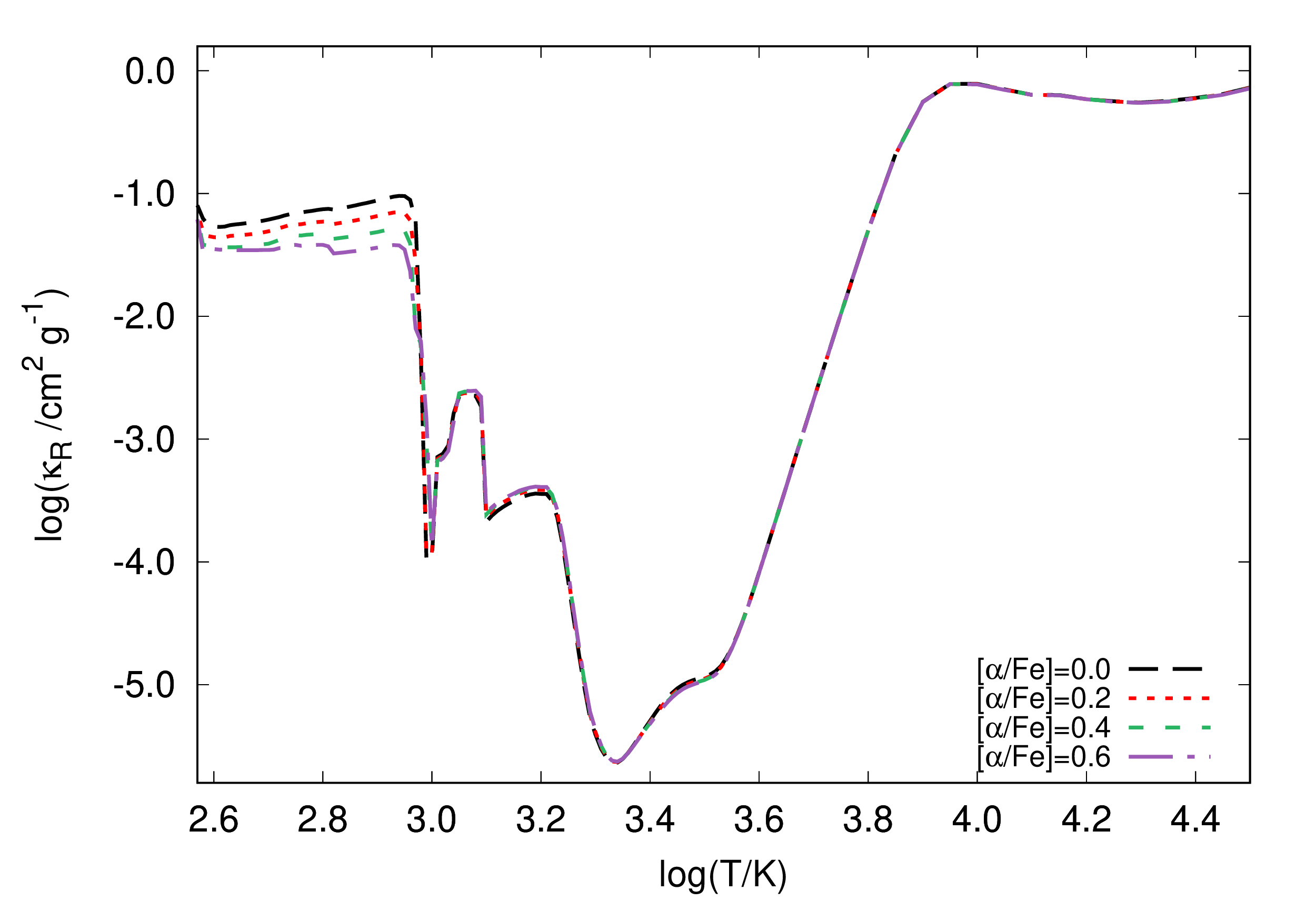}
    \caption{Properties of the Rosseland mean opacity as a function of chemical composition.  The reference chemical mixture is defined by $X = 0.7$ , with scaled-solar elemental abundances following \cite{Magg_etal_22}. 
     Left panel: Rosseland mean opacity for several values of metallicity. We take $\log(R)=-3$.
     Right panel: Rosseland mean opacity for $Z=0.001$ and a few degrees of alpha-enhancement, $\mathrm{[\alpha/Fe]}$. To maintain total metallicity, the increase in alpha-elements is balanced by a decrease in iron-group elements. We take $\log(R)=-5$.}
    \label{fig_comp}
\end{figure}

Figure~\ref{fig_comp} (left panel) illustrates the impact of metallicity changes on the Rosseland mean opacity. A scaled-solar composition is assumed in all cases except $Z=0$, where no metals exist.
As already noted by \cite{Ferguson_etal_05}, not only does the total opacity decrease as the amount of metals reduces, but condensation temperatures decrease as well, as fewer metals are available for the grains to exist in thermal equilibrium with.
We also explore the effect on the opacity caused by chemical mixtures with various levels of alpha-enhancement, focusing especially  on the temperature range where solid grains form (right panel of Figure~\ref{fig_comp}).
At  constant metallicity, there are no dramatic changes in opacity, except for two temperature ranges. As $\mathrm{[\alpha/Fe]}$ increases, \kR\ decreases for $300 \lesssim T/{\rm K} \lesssim 1000$, which is primarily due to a reduction in the abundance and hence opacity contribution of solid Iron (see Figures~\ref{fig_cond} and \ref{fig_kgrains}).
In a narrow temperature range around 125 K, we see a different pattern: the higher $\mathrm{[\alpha/Fe]}$, the greater the opacity. The increasing abundance of Gehlenite (Ca$_2$Al$_2$SiO$_7$), which is composed of a few alpha-elements, explains this increment. 

\section{Rosseland Mean Opacity Tables with Solid Grains}
\label{sec_tables}
We computed a standard set of opacity tables ranging from
$2.58 \leq \log(T/{\rm K}) \leq  4.5$ in 0.05 dex increments above $\log(T/{\rm K})=3.7$,  while the temperature resolution is increased in 0.01 dex increments  for $\log(T/{\rm K})\le3.7$. This enables more accurate tabulation in regimes where opacity may undergo abrupt changes and discontinuities due to the dominating contribution of negative hydrogen ions, molecules, and solid grains. For the density parameter {\em R} we cover a range $-8 \leq \log(R) \leq 1$ in steps of 0.5 dex.   Each table contains 129 temperature and 19 density points for a total of 2451 opacity values.

Opacity tables for other chemical compositions, or various choices of grain size, porosity and shape, can be easily computed upon request. It should be noted that the effects of changing the size distribution of dust grains, as well as their porosity and the presence of conglomerates of several dust species, have already been investigated by \citet{Ferguson_etal_07}. In particular, they find that ``changing the distribution of grain size has a marginal effect on the total mean opacity''. However, our investigation into the opacities for proplyds does not support this assertion.

\section{Concluding Remarks}
\label{sec_concl}
We compute the equation of state and provide Rosseland mean opacity tables for temperatures ranging from 30000 K to 400 K with the inclusion of the opacity of several dust species, coupling the \texttt{\AE SOPUS 2.0} and \texttt{GGCHEM} codes.

Due to the computational time cost, especially at low temperatures where dust grains can form, we could not implement the real time generation of opacity tables in \texttt{\AE SOPUS~2.0} web interface. Such an option is limited to the $3.2 \le \log(T/\mathrm{K}) \le 4.5$ temperature interval as described in \citet{aesopus2}.
Nonetheless users can retrieve Rosseland mean opacity tables extended to lower temperatures and optical constants of dust species from  the repository at \url{http://stev.oapd.inaf.it/aesopus\_2.0/tables}; a copy of these files have also been deposited to Zenodo: \url{https://doi.org/10.5281/zenodo.8221362}.

The chemical distribution of the abundances follows a scaled-solar pattern, according to a few relevant solar mixtures in the literature.
Below temperatures of $\simeq 1500$ K we include the opacity contribution of 43 solid grains. The grains are assumed to be homogeneous spheres, but other size, porosity and shape options may be considered in a future work and on user demand.
Additionally, opacity tables for other underlying chemical compositions may be added to the repository based on user requirements.
As an example, we present the opacities computed for a distribution with larger grains, more appropriate for the modeling of proto-planetary disks.

We recognize that in specific astrophysical environments, for instance for stars with powerful winds
\citep[asymptotic giant branch stars, red supergiants;][]{Hoefner_Olofsson_18, Ferrarotti_Gail_06}
dust formation does not happen in equilibrium as assumed here.
Also dust settling  is a critical process in proto-planetary disks \citep{Woitke_etal_16}.
In a follow-up work we may address these aspects.

The opacity tables computed in this work are not suitable for very low-mass stars, brown dwarfs and planets since the densities and gas pressures involved are typically too low for these cool objects.
In this perspective we plan to expand the opacity tables at higher densities (with $\log(R) > 1$), where electron degeneracy and other non-ideal effects, such as ionization potential depression, appear. Line pressure broadening should be considered in these conditions. This is crucial for modeling very low-mass stars, brown dwarfs and planets.

\begin{acknowledgments}
This research is funded by the Italian Ministerial Grant PRIN 2022, “Radiative opacities for astrophysical applications”, no. 2022NEXMP8, CUP C53D23001220006.
We also acknowledge support from Padova University through the research project PRD 2021. P.W. acknowledges funding from the European Union H2020-MSCA-ITN-2019 under Grant Agreement no. 860470 (\texttt{CHAMELEON}).
\end{acknowledgments}

\software{\texttt{\AE SOPUS} \citep{aesopus2, Marigo_Aringer_09},  
    \texttt{ExoCross} \citep{EXOCROSS_2018}, \texttt{DIANA} \citep{Woitke_etal_16},
    \texttt{GGchem} \citep{ggchem_18},
    \texttt{Opacity Project} \citep{Seaton_etal_94}.   
          }

\bibliography{opacdust.bib}{}
\bibliographystyle{aasjournal}

\end{document}